\documentstyle[aasms4]{article}

\def\cosmo{H$_0=50$~km/s/Mpc and q$_0=0\;$ }
\def\sec{^{\prime\prime\!\!}}

\begin{document} 

\input{psfig.tex}
\baselineskip=12pt
\title{\bf The HST Survey of BL Lac Objects: 
Gravitational Lens Candidates and Other Unusual Sources}
 
\author{Riccardo Scarpa\altaffilmark{1}, C. Megan Urry} 
\affil{Space Telescope Science Institute}
\authoraddr{Space Telescope Science Institute, 3700 San Martin Dr., 
Baltimore, MD 21218, USA}
\altaffiltext{1}{also at Department of Astronomy, Padova University, Vicolo 
dell'Osservatorio 5, 35122 Padova, Italy}
\authoremail{scarpa@stsci.edu,cmu@stsci.edu}

\author{Renato Falomo} 
\affil{Astronomical Observatory of Padova}
\authoraddr{Osservatorio Astronomico di Padova, Vicolo dell'Osservatorio 5, 
35122 Padova, Italy}
\authoremail{falomo@astrpd.pd.astro.it}

\author{Joseph E. Pesce} 
\affil{Eureka Scientific}
\authoraddr{ }
\authoremail{pesce@astro.psu.edu}

\author{Rachel Webster, Matthew O'Dowd}
\affil{Melbourne University}
\authoraddr{ Parkville, Victoria, Australia, 3052 }
\authoremail{ rwebster@isis.ph.unimelb.edu.au, modowd@isis.ph.unimelb.edu.au}
\and
 
\author{Aldo Treves} 
\affil{University of Insubria, Como, Italy}
\authoraddr{University of Insubria, Via Lucini 3, Como, Italy}
\authoremail{treves@uni.mi.astro.it}

\begin{abstract}

We present HST observations of seven unusual objects from the HST
``snapshot survey'' of BL Lac objects, of which four are gravitational
lens candidates. In three cases a double point sources is
observed: 0033+595, with 1.58~arcsec separation, and 0502+675
and 1440+122, each with $\sim 0.3$~arcsec separation. The last two also
show one or more galaxies, which could be either host or lensing
galaxies. If any are confirmed as lenses, these BL Lac objects are
excellent candidates for measuring H$_0$ via gravitational time delay
because of their characteristic rapid, high amplitude variability. An
additional advantage is that, like other blazars, they are likely
superluminal radio sources, in which case the source plane is mapped
out over a period of years, providing strong additional constraints on
the lensing mass distribution. The fourth gravitational lens
candidate is 1517+656, which is surrounded by three arclets forming an
almost perfect ring of radius 2.4~arcsec. If this is indeed an
Einstein ring, it is most likely a background source gravitationally
lensed by the BL Lac object host galaxy and possibly a surrounding
group or cluster. In the extreme case that all four candidates are
true lenses, the derived frequency of gravitational lensing in this BL
Lac sample would be an order of magnitude higher than in comparable quasar
samples.

We also report on three other remarkable BL Lac objects: 0138--097, which
is surrounded by a large number of close companion galaxies;
0806+524, whose host galaxy contains an uncommon arc-like structure;
and 1959+650, which is hosted by a gas rich elliptical galaxy with a
prominent dust lane of $\sim 5\times 10^5$ M$_\odot$.

{\underline{\em Subject Headings:}}
BL~Lacertae objects: individual (0033+595, 0138--097, 0502+675,
0806+524, 1440+122, 1517+656, 1959+650) ---
galaxies: structure --- galaxies: elliptical --- gravitational lensing

\end{abstract}

\section{Introduction.}

The HST ``snapshot survey'' of BL Lac objects\footnote{Based 
on observations made with the NASA/ESA
Hubble Space Telescope, obtained at the Space Telescope Science
Institute, which is operated by the Association of Universities for
Research in Astronomy, Inc., under NASA contract NAS~5-26555.} 
produced high resolution
images of $\sim100$ BL Lacs from six complete samples spanning the
redshift range $0.05 \lesssim z \lesssim 1.2$ (Falomo et al. 1998;
Urry et al. 1999b; Scarpa et al. 1999).
The main goal was to study the host galaxies and
near environments, and their evolution over cosmic time. In general,
the BL Lac objects lie in luminous elliptical galaxies, often
surrounded by groups or poor clusters, as has been found previously
from ground-based surveys (Wurtz, Stocke \& Yee 1996;
Pesce, Falomo \& Treves 1995). The excellent spatial
resolution of the HST WFPC2 allowed for better determination of galaxy
properties like morphology or core radius, and even in the relatively
short snapshot exposures, allowed easy detection of host galaxies out
to redshifts $z\sim0.5$.

HST WFPC2 spatial resolution also revealed new and unusual
morphologies in a handful of BL Lac objects. In this paper we report
seven unusual cases: three BL Lacs with double nuclei; a ring of 3
arcs surrounding a BL Lac; a BL Lac host galaxy with an isolated arc;
a BL Lac with many close companions; and a host galaxy with a
prominent dust lane. The double nuclei and the ring are new
candidates for gravitational lensing. Section~2 briefly reviews the
observations and data analysis, which are described more fully
elsewhere. In \S~3 we discuss individual objects, and in \S~4 we give our
conclusions. Throughout the paper \cosmo are used.

\section{Observations and Data Analysis}

\subsection{WFPC2 data} 

Observations and data analysis are described fully by Falomo et al. (1997),
Urry et al. (1999a), and Scarpa et al. (1999) and are only briefly 
reviewed here. All BL Lac objects were observed
with the Wide Field and Planetary Camera 2 (WFPC2) through the F702W
filter. Targets were centered on the PC chip, which has pixels
$0.046$~arcsec wide. To obtain a final image well
exposed in both the inner, bright nucleus and in the outer regions
where the host galaxy emission is still above the wings of the
Point Spread Function (PSF), we
made a series of increasingly longer exposures with total 
duration ranging from 300 to
1000 seconds. The journal of the observations in Table~1 gives the
coordinates (J2000), date of observation, total exposure time in seconds, 
and reported redshift of the target BL Lac objects.

After preliminary reduction carried out as part of the standard HST 
pipeline processing (flat--fielding, dark and bias subtraction, 
and flux calibration), we simultaneously combined images and 
removed cosmic rays using the IRAF task ``CRREJ.'' 
Fluxes were converted to R-band magnitudes following the
prescription of Holtzman et al. (1995, their Equation~9 and Table~10).
Finally, we modeled the PSF in two parts: 
the core using the Tiny Tim software (Krist 1995), 
and the wings (at $\gtrsim 2$~arcsec radius) using the average 
of well-exposed stellar images (Urry et al. 1999a; Scarpa et al. 1999).

\subsection{NICMOS data}

An HST NICMOS observation of the BL Lac object 0502+675 was carried out
as part of a related but separate survey of BL Lac objects. The
observation was on May, 5, 1998, through filter F160W, which is
equivalent to the standard H band, with the NICMOS
camera 2, which has pixel size of $0\sec .075$. Due to the lower resolution
of HST in the infrared, this pixel size offers a sampling of the PSF as
good as the PC camera in the R band. Three separate images were
obtained, dithered among three positions in order to
better estimate the contribution of the sky to the total signal. The
data were first reduced and flux calibrated in the standard HST
pipeline, then the effect of the random bias (known as
the ``pedestal'') was removed. Images were cleaned from cosmic rays 
and other defects,
and a sky frame, obtained by median filtering the three frames, was
subtracted. Each image was then re-sampled, increasing the sampling
by a factor of 2, and finally re-centered and combined.

\section{Results and Discussion of Individual Objects}

\subsection{1ES~0033+595}

The HST image of this BL Lac object, an Einstein Slew Survey source
(Perlman et al. 1996), shows two objects of similar brightness at the
reported optical position, separated by 1.58~arcsec (Fig. 1). Neither
the VLA radio map nor the optical finding chart given by Perlman et
al. (1996) indicate the source is double, nor would they given their
low spatial resolution.

The two sources ``A'' and ``B'' have magnitudes $m_{R} =
17.95\pm0.05$~mag and $m_{R} = 18.30\pm0.05$~mag, respectively.
Object ``B'' is at position angle 63$^\circ$ with respect to ``A.''
Absolute coordinates for both components, derived using the HST
astrometric solution, are given in Table~1. A faint, clearly resolved
object (``G'') is also detected south of the two brighter
objects. Its radial profile is consistent with that of an elliptical
galaxy of total integrated magnitude $m_{R} = 22.3\pm 0.15$~mag.
The galaxy is
$1.\sec 39$ south of source ``B'' and $1.\sec 43$ SE of source ``A.''
There is no reported redshift either for the BL Lac object or for galaxy ``G.''

Both ``A'' and ``B'' appear to be unresolved, although evaluating
their radial profiles is made difficult by the small separation. In
order to extract the radial profile of each source, we first modeled
and subtracted the companion. As a model, we used a Tiny Tim
generated PSF (Krist 1995), computed with a factor of 3 
oversampling. The PSF was centered on the object with precision $\sim
0.05$ pixels, then re-sampled and convolved with the PC camera
kernel. The flux normalization was done matching the source flux in an
annulus with $2 < r < 5$~pixels (avoiding the central pixel which
could be saturated). The radial profiles of the two sources and the
PSF are very similar, with only minor deviations that are well within
the uncertainties (not reflected in the error bars) introduced by
the closeness of the two sources (Figure~2). The fact that both
profiles are slightly above the PSF may be due simply to a defective
subtraction of the nearby companion. Based on the
similarity of the two source profiles and their consistency with the
PSF model, we conclude that both ``A'' and ``B'' are unresolved.

Two point-like images can be produced by a chance alignment of the BL
Lac with a foreground star or with a foreground or background AGN; or
because the nucleus is physically double, as is sometimes observed in
quasars (Kochanek, Falco \& Mu\~noz 1997); or because of
gravitational lensing of a background point source by a foreground
mass. The easiest way to shed light on this issue is uniquely
identifying one of the two sources with the radio counterpart. The
precision of the coordinates of the only available VLA map is $\sim
1$~arcsec (Perlman et al. 1996), and so at present it is not possible
to discriminate which (if only one) of the two optical point sources
is the radio source. Formally the
radio position agrees better with the HST coordinates for the eastern
(``B'') source ($\Delta \alpha = 0\sec .0$ and $\Delta \delta = 0\sec .1$)
than with the western source (``A''; $\Delta \alpha = 0\sec .2$ and
$\Delta \delta = 0\sec .6$). Interestingly, the VLA radio map does show
a slight elongation exactly along the line connecting the two optical
images, suggesting there may be two barely-resolved radio sources
coincident with the two optical images.

Given the relatively large separation of the two sources and the low
galactic latitude of 0033+595, the
probability of a chance alignment of the BL Lac with a foreground star
is not small. 
We derive the stellar surface density toward 0033+595 directly from the HST
image. There are 33 stars as bright as object ``B'' in the whole WFPC2 field of
view ($\sim20550$ $arcsec^2$). This gives a probability 
of $\sim1$\% of having a star
within a ring of radius $1.\sec 58$ from ``A,'' which for a 
sample of 100 targets means that we actually expect to find 1 case 
like this (note however that not all 100 observed BL Lac objects 
are at this low galactic latitude).

That object ``A'' is possibly a foreground star is also
indicated by the much bluer color of object ``B'' ( U-B=0.4~mag and
$-0.1$~mag for ``A'' and ``B,'' respectively; Falomo \& Kotilainen,
private communication).
Based on radio coordinates and color, then, it
seems quite possible that ``B'' is the true BL Lac and ``A'' is a star. If
this is the case, then the radio map is centered on ``B'' and the
elongation is on the opposite side with respect to ``A,'' and is not
associated with it. In Figure 3 we show a finding chart for this
object with source ``B'' marked as the most probable BL Lac.

However, the hypothesis of gravitational lensing
can not be excluded since the different color 
could be due to differential internal absorption through the
lens. To produce a $\Delta$U--B of 0.5~mag,
source ``A'' should be reddened by A$_V\sim2$~mag
more than ``B,'' corresponding to a moderate hydrogen column density
of $3.3\times 10^{21}$~cm$^{-2}$.

The presence of significant absorption is consistent with the lack
of a strong blue continuum observed in the only published spectrum of
0033+595 (Perlman et al.1996). This spectrum was obtained with a
$2\sec .5$-wide slit that included both sources; it is featureless, and
intriguingly red for a BL Lac object. If ``A'' is a red star, strong
absorption lines should be present in the spectrum but they are not
seen. Similarly, if it were an unrelated quasar, one might expect to
see emission lines. The absence of such spectral features does not
rule out these possibilities but makes them less likely, given the
approximately equal brightness of the two point sources.

Spatially resolved spectra of each point source are crucial, and
under good seeing conditions and with a properly oriented slit, should
be easy to obtain from the ground. If the two point sources turn out to have
identical spectra, the unusually red spectrum may be produced by
internal absorption in an otherwise undetected, aligned lens.

\subsection{1ES~0502+675}

This Einstein Slew Survey source was identified as a BL Lac object on
the basis of its featureless optical spectrum (Schachter et al. 1993).
At HST resolution 0502+675 is clearly double, with separation of only
0.33~arcsec (Fig. 4). 

A careful comparison of the two radial profiles,
(extracted as described for 0033+595), shows that the brighter object
is more extended than the fainter one and differs significantly
from the PSF profile (Fig.~5).
We therefore fit the radial profile of the brighter source with a PSF
plus elliptical galaxy (de Vaucouleurs) model, convolved with the
PSF. The best fit yields a point source of m$_R = 17.3\pm0.1$~mag 
and a surrounding galaxy of total magnitude m$_R = 18.9\pm0.1$~mag 
(integrated to infinity) and half-light radius
$r_e=0.6 \pm 0.07$~arcsec. The addition of the galaxy is significant at the
99.99\% confidence level according to an F-test. A disk galaxy can
not be ruled out, but given that BL Lac hosts and gravitational lenses
are overwhelmingly elliptical galaxies, it is most
likely to be this morphological
type. Using these parameters we subtracted a two-dimensional model
from the brightest object, and re-extracted the radial profile of the
companion; it is consistent within the errors with the PSF profile
(Fig. 5), and the derived magnitude of the point source is m$_{R} =
18.7\pm0.2$~mag.

Because of the small spatial separation, the likelihood that this
close binary is the result of chance superposition is much less than
in the case of 0033+595, even under the conservative assumption that
the BL Lac is the brighter object and the fainter object is an
unrelated source. At high galactic latitude
there are $3\times10^3$ stars per square degree
brighter than m$_{R} = 19.0$~mag (Bahcall \& Soneira 1980), and a
corresponding probability of $P\lesssim 7\times10^{-5}$ for a
faint foreground star being as close as $0\sec .33$ to the BL Lac object, or
$7\times10^{-3}$ for a sample of 100 objects. The probability for a
chance alignment is therefore rather small.

A possible explanation for the presence of an object so close to the BL
Lac is that it is a companion galaxy, frequently observed near BL
Lacs. However, even at HST resolution the alleged companion is
unresolved and the difference in luminosity between the two objects,
$\Delta m_R = 1.4$~mag, is much less than the typical difference of
several magnitudes between BL Lac objects and companion galaxies
(Pesce, Falomo \& Treves 1995; Falomo 1996), arguing against the suggestion
that they are companion objects. The possibility of a double nucleus
is somewhat more likely, as the two point sources have similar
magnitudes, but in this case the galaxy is unlikely to be centered on
one of the two, as observed.
 
Fortunately, as part of a related but separate snapshot survey of BL
Lac objects carried out with the NICMOS camera 2, 0502+675 was
re-observed in the H-band. Both sources were clearly detected, as well as
the galaxy surrounding them (Figure 4).
The two point sources have, within the errors, the same luminosity 
ratio in both bands, i.e., they have the same R-H colors.
For the brighter and fainter sources we measure $m_H=15.4\pm0.1$~mag and
$16.9\pm0.2$~mag, respectively. The corresponding optical-IR spectral
index is $\alpha = -0.7$, typical for a BL Lac. 

The best fit of the galaxy radial profile with a de Vaucouleurs 
model gives $m_H=15.7\pm0.2$~mag and $r_e=0.5\pm0.2$~arcsec; the
latter agrees very well with the value derived from the R-band image.
Two-dimensional fitting of the outer galaxy isophotes 
is severely hampered by the presence of the
two strong point sources, so we were not able to determine whether the
galaxy is actually centered on one of the two point sources or
between them.

Recently a redshift of $z=0.314$ has been reported, 
based on detection of CaII H\&K and MgII
absorption lines (Perlman, private communication).
At this redshift, the K-corrected absolute magnitude of the galaxy is
$M_R=-23.2$~mag (including a K-correction of 0.4~mag in the R band while
in the H band the K-correction is negligible) and $M_H=-26.0$~mag. The
corresponding color is $R-H=2.8$~mag, to be compared with 2.9~mag expected for
an early type galaxy at that redshift (Kotilainen, Falomo \& Scarpa
1998 and references therein). 
However, the effective radius at that redshift (3.1~kpc)
is relatively small for a galaxy of that absolute magnitude, and is
smaller than the average value for resolved host galaxies in the full
survey of 100 BL Lac objects, $\sim 10$~kpc (Urry et al. 1999b). 
Rather than being an unusually compact host, this galaxy could
instead be a closer, intervening elliptical.

Given the wide separation in wavelength between the two HST observations, 
the similarity of the two point source spectra 
strongly favors the gravitational lensing hypothesis. 
To prove this is actually a lens requires
spectroscopy with very high spatial resolution, which will be done in
Cycle~8 with the HST STIS long-slit (PI Scarpa). 
If the lensing scenario is confirmed,
the observed galaxy is a good candidate for the lensing mass. 

\subsection{1ES~1440+122}

This is another Einstein Slew Survey source, identified as a BL Lac
object on the basis of a nearly featureless spectrum, with only weak
Ca II H and K break at $z=0.162$ (Schachter et al. 1993). In the HST
PC image (Fig.~6), 1440+122 is double, consisting of a large
elliptical galaxy with a bright central point source (``A''), and a
second point source just 0.29~arcsec to the east (``B''). A second
galaxy, without a central point source, lies 2.53~arcsec to the west
(``G''). The radio and optical positions reported for 1ES~1440+122
(Perlman et al. 1996) coincide most closely with ``A,'' which is most
likely the BL Lac object, while ``G'' is either an unrelated galaxy or
an unusually bright companion galaxy.

Since the identification spectrum was obtained through a
large aperture, it likely included light from both ``A'' and ``G,''
meaning the reported redshift could refer to either object. The
signal-to-noise ratio is also rather low, and all absolute quantities
should be considered with caution.

We fitted object ``A'' with a model of an elliptical galaxy plus point
source (Fig.~7), convolved with the PSF, and obtained best-fit values of the
galaxy magnitude (integrated to infinity), m$_R = 16.70\pm0.05$~mag;
half-light radius, $r_e = 3.9\pm0.25$~arcsec; and point source
magnitude, m$_R=16.9\pm0.1$~mag. For redshift $z=0.162$, these correspond
to absolute magnitude $M_R=-23.6$~mag (including a K-correction of
0.2~mag), r$_e=15$~kpc, and point source luminosity M$_R=-23.2$~mag.
Similar results for the host galaxy were reported by Heidt et al. (1999).

The companion source ``B'' has magnitude m$_R=19.8$~mag and is 
located 0.29~arcsec east of ``A'' at position angle 70$^\circ$. 
Its radial profile is consistent with the PSF but 
the closeness of the brighter source ``A'' makes it 
difficult to determine whether it is unresolved.

The nearby galaxy ``G'' (at position angle $\sim 260^\circ$) also has 
an elliptical morphology, with total apparent magnitude m$_R = 17.56$~mag
and $r_e = 2.8$~arcsec, which at $z=0.162$ would correspond to 
M$_R = -22.7$~mag (including K-correction as for ``A'')
and $r_e = 10.6$~kpc. 
If the redshift is correct for both galaxies, they would be
large, bright ellipticals, similar to typical BL Lac host galaxies.

A chance superposition of ``A'' and a foreground star (``B'') is
possible, and somewhat more likely (P$\sim 1\%$) than for 0502+675 due
to the faintness of ``B,'' but less likely than for 0033+595 because
of the small angular separation of the two point sources. A double
nucleus is highly unlikely given that the galaxy is centered on one of
the point sources and they have very different magnitudes.

It remains possible that ``B'' and the point source in ``A'' are two images
of a distant background blazar, and that the galaxy in ``A'' is actually the
foreground lensing galaxy. With such small splitting, the fact that
that the galaxy appears well centered on one of the two point sources
(to within 0.05~arcsec) is not unexpected.

Again, confirmation requires spatially resolved spectroscopy, 
which will also be done in Cycle~8 with the HST STIS. We note
that galaxy ``G'' will fortuitously fall within the STIS long slit.

If 1440+122 is not lensed, and the point source and galaxy in ``A'' are 
physically associated (i.e., have the same redshift), this system is still
very interesting. Specifically, if galaxy ``G'' is at the same redshift,
it would be a relatively rare occurrence of two luminous companion
galaxies so close together, only $\sim10$~kpc projected distance at $z=0.162$.
While BL Lac objects are often found in groups or poor clusters, 
most companions are much smaller and less luminous than the BL Lac host
galaxy (e.g., Falomo 1996), suggesting the BL Lac dominates the system.
This is in contrast with what is observed for radio galaxies, which are more
often found in such dumbbell systems (e.g., Fasano, Falomo \&
Scarpa 1996 and references therein). Detection of more cases like this
would strengthen the ``unification'' connection between BL Lac objects
and radio galaxies.

It is worth noting that a dumbbell system like 1440+122 (or the one of
1415+529, Wurtz, Stocke \& Yee 1996), when observed from the ground
under normal seeing conditions, would be unresolved or only marginally resolved
(especially if it were at larger redshift). In that case the central
point source would also appear off-center with respect to the host. As
an example, the BL Lac object MS~0205+315 was reported to have an
off-center disk host (Wurtz, Stocke \& Yee 1996) but in better seeing
was found to have a normal elliptical host and a large companion
galaxy (Falomo et al. 1997).

\subsection{H~1517+656}

The HEAO1-A3 BL Lac 1517+656 (Elvis et al. 1992) is perhaps the most
unusual object discussed here. At HST resolution the bright,
unresolved BL Lac nucleus (m$_R=16.2$~mag) is surrounded by three arcs
describing an almost perfect ring of radius 2.4~arcsec (Fig.~8). The
ring is off-center by $\sim 0.5$~arcsec with respect to the BL Lac but
otherwise resembles an Einstein ring. Given the low signal-to-noise
ratio in this image, it is also possible that the arclets describe two
different rings centered on the BL Lac, one traced by the 2 innermost
arclets, another by the outermost one. The surface brightnesses of
the arcs are approximately the same, $\mu_{R} \sim
22.4$~mag~arcsec$^{-2}$, and they are resolved radially, having a
width of $\sim 0.2$~arcsec. The two bright, resolved spots at
position angles 126$^\circ$ and 260$^\circ$ have magnitudes $m_{R} =
23.6$~mag and 23.8~mag, respectively.

Due to the shortness of the exposure, 320 seconds, the arcs, 
while clearly detected, are severely under-exposed and the signal-to-noise
ratio is low. In a deeper image, the ring could well be filled in with
fainter structures. Note that the narrowness of these arcs makes them 
very difficult to detect from the ground where, even in excellent seeing, 
they would be have surface brightness lower by $\sim2$-4~mag~arcsec$^{-2}$, 
as well as greater contamination by scattered light from the BL Lac nucleus.
(Our attempt to image the arcs using the much larger Canada-France-Hawaii 
Telescope was not successful, even with 10 times the exposure time.)

MgII and FeII absorption lines place a lower limit to the BL Lac 
redshift of $z>0.7$ (Beckmann et al., private communication). 
The lack of an exact redshift adds uncertainty to the interpretation 
of the arcs. 
We discuss three scenarios:

\begin{itemize}

\item{1.} The most likely option is that one or more background galaxies are
gravitationally lensed to give the observed arclets. 
In this case the BL Lac would necessarily be in the foreground; 
otherwise, if it were the central image of the lens, 
it would be heavily de-magnified and a second image should be 
observed outside the ring. 
If the BL Lac is in the foreground, it is contributing to
the lens (since the almost perfect alignment of more than two objects 
is highly unlikely). BL Lac objects are often associated with other galaxies,
and these could account for the center-of-mass being offset from the
BL Lac and for the large diameter of the ring, 
$4.8$~arcsec, which corresponds to a lens mass of a few times 
$10^{12} M_\odot$ depending on source/lens redshifts and cosmology.
These additional galaxies should be
visible in a deeper HST image and may already have been seen in
the two resolved spots. 

In order to establish whether these arcs indeed constitute an Einstein
ring, deeper images with $\sim0.1$~arcsec resolution are required.
Multicolor images would also constrain the
redshift of the arcs, if indeed they represent stellar light.
If confirmed, this would be the first discovery of a BL Lac object acting
as a lens rather than being lensed itself.

\item{2.} The distant BL Lac object is located behind a foreground face-on
spiral galaxy, possibly an Sc with a very small nucleus. Due to the
short exposure time, we observe only the brightest regions of the
spiral arms. The BL Lac may be sufficiently off-set from the galaxy
nucleus so that it is not macro-imaged, though it could still be
micro-lensed by stars in the galaxy.

The primary argument against this picture
is the high surface brightness of the arc-like features.
The central surface brightnesses 
of nearby spiral disks are roughly constant, $\mu_R= 
20.5$~mag~arcsec$^{-2}$ (Freeman 1970; Van der Kruit 1987,1989),
not much higher than the surface brightnesses of the observed arcs,
which are well away from the putative galaxy center. 
Hence even without the $(1+z)^4$ cosmological dimming, 
the ring is already much brighter than expected for a spiral galaxy.
In addition, the spectrum of 1517+656 
is not unusually red (Beckmann et al., private communication), 
in contrast to the one other BL Lac thought to be located behind a spiral 
galaxy (1413+135; Carilli, Perlman \& Stocke 1992).

If this hypothetical galaxy were responsible for the detected
MgII and FeII absorption lines at $z\sim0.7$, its diameter 
would be $\gtrsim 45$~kpc, unusually large for a spiral galaxy
though not impossible. (Of course, if the galaxy is closer, 
this consideration does not apply.) 

We conclude that this scenario is quite improbable.

\item{3.} The arcs are part of a galaxy associated with the BL 
Lac nucleus, either
bright arms of a spiral, or shells surrounding an elliptical. In the
former case, 1517+656 would be the first BL Lac object for which
spiral arms are unambiguously detected. (The possibility of spiral host
galaxies in 2-3 other BL Lac objects remains extremely controversial;
see discussion in Urry \& Padovani 1995, and Scarpa et al. 1999.) 
However, the arguments
about size and surface brightness make this possibility doubtful.

If the arcs are elliptical shells, subject to $(1+z)^4$
cosmological dimming, the intrinsic surface brightness would be
$\mu_{R}\lesssim 20$~mag~arcsec$^{-2}$, at least several magnitudes brighter 
than usual (Forbes \& Reitzel 1995). Moreover, such shells always have
surface brightnesses much fainter than the galaxy producing them
(Malin \& Carter 1983), and so we should clearly see the galaxy, 
whereas the BL Lac nucleus is unresolved.

\end{itemize}

Of these three possibilities, the first is the most plausible.
To establish whether these arcs indeed constitute an Einstein ring
requires deeper images with 0.1-arcsec resolution. 

\subsection{PKS~0138--097}

This BL Lac object from the 1~Jy sample (Stickel et al. 1991) is 
located in a rich environment, 
has a smooth IR-optical spectrum (Fricke et al. 1983), and is highly
polarized (Impey \& Tapia 1988).
Weak emission lines from MgII, [OII] and [NeV] were recently detected at
$z=0.733$ (Stocke \& Rector 1997), together with a previously known 
absorption system at $z=0.501$ (Stickel et al. 1993).
The BL Lac object appears bright and point-like in our HST image (Fig.~9). 
To increase the signal-to-noise ratio and better investigate whether it is
actually resolved, we extracted the azimuthally averaged radial profile,
masking out nearby sources. The comparison of the radial profile of the 
BL Lac with the PSF model shows a small but systematic departure beyond
1~arcsec (Fig.~9), in excess of our estimated 
statistical uncertainties.
However, after increasing the sky of $1\sigma$, the radial profile is
fully consistent with the adopted PSF model and we conservatively
conclude that the source remains unresolved.
At the 95\% confidence level, the upper limit to the host magnitude is 
m$_R>20.1$~mag, or M$_R>-25.4$~mag at $z=0.733$.

The environment near 0138--097 is rich, with at least four galaxies
detected within a radius of 3~arcsec, corresponding to a projected
distance of 10~kpc (Heidt et al. 1996, whose designation we adopt
here). In our HST snapshot image (Fig.~9) three of the four are
clearly visible, while object ``A'' is not detected, as expected given
the short snapshot exposure time. The apparent R magnitudes, together
with the distance from the BL Lac, are given in Table~2 for each
galaxy. Our measurements are generally in agreement with those of
Heidt et al. (1996), apart from object ``D'' which we estimate is
approximately one magnitude fainter, and object ``E,'' for which 
they reported no data.
Source ``C'' is clearly resolved, with major axis at
position angle $\sim 90^\circ$. The projected distance is only 14~kpc at
$z=0.733$, so there could well be gravitational interaction with
the BL Lac.

The presence of absorption systems and close companions 
led Heidt et al. (1996) to suggest this BL Lac object may be 
affected by gravitational micro-lensing,
Galaxy ``C'' is a good candidate for producing the absorption line,
given the small projected distance from the BL Lac object,
and possibly for micro-lensing as well.

\subsection{1ES~0806+524}

This BL Lac object from the Einstein Slew Survey (Schachter et al. 1993)
has a rather interesting morphology. In our WFPC2 F720W image (Fig. 11), 
the BL Lac nucleus is surrounded by a bright elliptical galaxy,
as it is typical of BL Lacs at low redshift. 
Not typical at all is the large arc-like structure 1.93~arcsec 
south of the nucleus. 

The radial profile of 0806+524 is well described by a point source
plus a de Vaucouleurs host galaxy (the contribution of the arc to the
azimuthally averaged light is negligible). 
The two components have comparable luminosity, with best-fit
values m$_R = 16.3$~mag for the central point source, and
m$_R = 16.7$~mag and $r_e = 1.7$~arcsec for the elliptical galaxy.

For this BL Lac there is only a tentative redshift.
The discovery optical spectrum was featureless (Schachter et al. 1993), 
as were two subsequent spectra of better quality (Bade et al. 1994; 
Perlman et al. 1996), but Perlman et al. (private communication) have since
obtained a new spectrum from which a redshift
$z=0.136$ was estimated from the Ca H\&K absorption line. 
At this redshift, the absolute magnitude of the host galaxy would be
M$_R=-23.25$~mag (including a K-correction of 0.2~mag),
quite normal for a BL Lac host galaxy (Urry et al. 1999a; 
Wurtz, Stocke \& Yee 1996; Falomo 1996).

The arc has radius of curvature of $\sim 2$~arcsec, 
is roughly centered on the BL Lac nucleus, and its surface
brightness is $\mu_R\sim 22.2$~mag~arcsec$^{-2}$. Although the nature
of the arc can not be determined with only one image, a possible
explanation is that it is a shell, not an uncommon feature in
ellipticals but uncommonly bright in this case (Malin \& Carter 1983, 
Forbes \& Reitzel 1995). If so, it would be the first detection of 
such a shell in a BL Lac host galaxy. This idea is supported by the
centering of the arc on the nucleus. 
The lack of a symmetric shell on the other side of the galaxy, can be due
to the faintness of the structure and/or to projection 
effects, especially if the detected arc lies in the foreground and the other
in the background with respect to the galaxy.
Alternatively, it could be the remnant of a past interaction, although
there are no other signs of tidal disturbance, nor is the nucleus offset
from the host galaxy.

\subsection{1ES~1959+650}

This Einstein Slew Survey source was identified as a BL Lac 
based on the high optical polarization and ratio of radio to X-ray emission 
(Schachter et al. 1993). The optical spectrum (Perlman et al. 1996) is
dominated by the starlight from the host galaxy, which is well
resolved in our HST image, with radial profile well described by a 
de Vaucouleurs law plus a point source;
a pure disk model is ruled out. Best-fit parameters are
m$_R=14.92\pm0.05$~mag and r$_e=5.1\pm0.1$~arcsec for the galaxy and 
m$_R=15.4\pm0.1$~mag for the point source. 
At redshift $z=0.048$, this corresponds to
M$_R=-22.5$~mag and r$_e=6.6$~kpc for the galaxy and 
M$_R=-21.5$~mag for the nucleus.
Heidt et al. (1999), fitting a PSF plus de Vaucouleur model, found similar
values, M$_R(host)=14.77$~mag and $r_e=11$~kpc.

Some small deviations
from the r$^{1/4}$ law are however evident, indicating
a disturbed morphology. Indeed, after subtracting a scaled PSF from
the image, a prominent dust lane is apparent along the major axis, 
$\sim0.2$~arcsec north of the nucleus (Fig. 12). 

With a single image, it is very
difficult to calculate the mass of gas in the dust lane. We make a
rough estimate assuming the galaxy is intrinsically symmetric and the
difference in luminosity between opposite sides is entirely due to 
dust absorption. The dust extinction is then $A_R = 0.26$~mag in the R band, 
or $A_V = 0.35$~mag in V using the extinction curve of Cardelli, Clayton \& 
Mathis (1989). This can be converted to hydrogen column density via
$E(B-V) = 3.1 A_V = \log N_H - 21.83$ atoms~cm$^{-2}$~mag$^{-1}$ 
(Shull \& Van Steenberg 1985; Rieke \& Lebofsky 1985).
Given the observed dust surface area of $\sim 6\times10^{42}$~cm$^2$,
the total gas mass is $\sim 5\times10^{7}$~M$_\odot$, and, finally
a dust mass of $\sim 5\times10^{5}$~M$_\odot$ for a gas-to-dust ratio
of 100 (Bohlin et al. 1978). 
This figure is not unusual in normal elliptical galaxies 
(Wiklind, Combes \& Henkel 1995), but 
has not been reported previously in a BL Lac host, possibly 
because of the lower resolution of ground-based images,
in which the dust lane could be washed out by scattered light from
the nucleus and/or host galaxy.

\section{Discussion and Conclusions}

We have presented HST observations of seven unusual BL Lac objects
from our HST snapshot survey (Urry et al. 1999b). Four are
tentative candidates for gravitational lensing. These include three
close doubles with one or more nearby galaxies that are plausible
foreground lensing galaxies. For 0033+595, given the non negligible
probability of a chance alignment with a foreground star, and the
clearly different colors of the two components, the possibility 
of a gravitational lens is not strong. Color and VLA coordinates
suggest the BL Lac is object ``B'' (Figs. 1 and 3). The case of
0502+675 is much stronger because the two images have similar
luminosity ratios in both R and H bands, $\Delta m = 1.4\pm0.2$~mag and
$1.5\pm0.2$~mag, respectively. The case of 1440+122 remains unclear due
to the lack of an image in a second filter. The lensing scenario can be tested
with spatially resolved spectroscopy; for the candidate with
1.6~arcsec spacing this can be done from the ground, while for the
other two, both of which have separations of $\sim0.3$~arcsec, HST or
comparable resolution is required.

The fourth candidate for gravitational lensing is a partial ring of
three arcs, possibly an Einstein ring, slightly offset from the unresolved 
BL Lac object. If this is a case of lensing, then
the BL Lac would be in the foreground, its host galaxy and (likely) 
surrounding cluster constituting the lensing mass, 
while the arcs would be images of one or more background galaxies. 
This is the opposite of the usual scenario, where the AGN is lensed by
a foreground object.
There are two resolved spots on the arcs themselves;
if these were galaxies associated with a cluster around the BL Lac object
(at $z>0.7$), then their absolute magnitudes would be 
M$_R \lesssim -22.7$~mag (including a K-correction of 1.5~mag).
This figure is not unreasonable for a galaxy, but is $\sim1$ magnitude
brighter than an M$^*$ galaxy, so it is possible that those structures
are actually background object(s) magnified by the lens.
Our WFPC2 image was not very deep, so additional galaxies within the ring 
and/or additional structure in the rings should easily be visible in 
a deeper HST image. (Because they are very thin, the rings would have
very low surface brightnesses in typical ground-based seeing, and so 
would be difficult to detect, and in any case they will remain unresolved,
losing important lens-mapping information.)
The alternative (non-lensing) explanations cannot be ruled out but
seem somewhat less likely. Whatever the case, this BL Lac object remains unique.

The redshift distribution of BL Lacs in our snapshot survey sample peaks below
$z\sim0.5$, although the redshifts of BL Lacs are poorly known due to
the weakness of detected features. About 1/3 have no measured
redshift, and many measured values are either lower limits based on
intervening absorption, as for 1517+656, or are based on features in the
``host'' galaxy. In at least some cases, like 1440+122, the reported
redshift could refer to the lensing galaxy rather than the BL Lac
object (Ostriker \& Vietri 1985). Hence, it is possible that
the average redshift of our sample is somewhat bigger than
$z\sim0.5$. At such a small redshift, the probability of
lensing is vanishingly small, so that if all lens candidates were
confirmed, the incidence of lensing in this sample of 100 BL Lac
objects would be much higher than in comparable surveys of quasars.
For comparison, in samples of quasars with $1\lesssim z \lesssim 2.5$,
the incidence of strong lensing is roughly 1-2 in 1000 (Kochanek
1996), albeit at lower spatial resolution,
and the expected number decreases rapidly with decreasing AGN
redshift.

Considering only lens candidates with separation $>1 arcsec$, we have only one
(weak) case, 0033+595. The HST snapshot survey for lenses found
$\lesssim 4$ lenses with such large separation in a sample of $\sim
500$ bright quasars (Maoz et al. 1993). Allowing for the smaller size
of our sample and a factor $\sim 2$ lower mean redshift, we would
expect $\ll 0.4$ lenses in our BL Lac sample. Thus, for separations
of $\sim1$~arcsec or more, our survey may not be terribly out of line,
especially considering the small number statistics.

Interestingly, Stocke and Rector (1997) reported an over-density, by a
factor of 4-5, in the number of MgII absorption systems detected in
the spectra of BL Lacs relative to quasars, and suggest this is the result
of a magnification bias due to micro-lensing. The magnification bias
for a particular sample depends on the steepness of the differential
number counts. The high frequency of absorption lines and the large number
of candidate lenses may both be due to a large magnification bias if
the luminosity function of BL Lacs is steep. Note that, out of
the $\sim 300$ BL Lacs known, at least one is already a confirmed lens
(0218+357, with separation $0.34$~arcsec; Patnaik et al. 1992, 1993). 

Time delays in gravitationally lensed, variable AGN have been used to
estimate the Hubble constant. For the three best cases, the ensuing
errors on $H_0$ remain large simply because of uncertainties in the
lens mass model: (1) 1115+080, for which H$_0= 44 \pm 4$ or $65 \pm 5$
km~s$^{-1}$~Mpc$^{-1}$ (Schechter et al. 1997, Impey et al. 1998); (2)
0957+561, for which the published value is H$_0=64 \pm
13$~km~s$^{-1}$~Mpc$^{-1}$, not including modeling uncertainties
(Kundic et al. 1997); and (3) the BL Lac object 0218+357, for which a very short
time delay, $12\pm 3$ days, gives H$_0 \sim 60$~km~s$^{-1}$~Mpc$^{-1}$
(Corbett et al. 1996). In the latter case, thanks to the large and rapid flux
variability, combined with the small separation of the BL Lac images,
only 80 days of observation were required to measure the time delay,
to be compared with the almost 20 years spent on 0957+561. Further,
the resolved milliarcsecond radio structure of 0218+257 greatly
constrains the lens mass model. Since BL Lac objects are
likely to be expanding superluminally, the source itself will map out
the lensing plane in just a few years!

The uncertainties in measuring $H_0$ in individual lenses argues that
new lenses are quite useful. Especially valuable are lensed BL Lacs,
in which the flux variability should be large and rapid (Ulrich,
Maraschi \& Urry 1997), and the superluminal expansion of the radio
source offers new and powerful constraints. Our close separation
pairs, 0502+675 and 1440+122, if confirmed as gravitational lenses,
will be especially powerful and convenient tools for estimating $H_0$.
The time delay calculated using a simple isothermal mass model and a
source redshift of $z\sim1$ is $\sim 35$~days. (More detailed models
can be determined when the source and galaxy redshifts are known.) In
this case, monitoring over a period of 4-6 months should be sufficient
to estimate a useful value of $H_0$.

If instead these BL Lac pairs are shown to be true binaries, they will
be the first such close-separation pairs of quasars, which can be used
to study the role of tidal interactions in the AGN phenomena. At
present, we do not have the data necessary to discriminate among
these different hypotheses. To be confirmed as gravitational lenses,
spectroscopic observations are required. 

Among the other unusual sources presented in this work, 0806+524 shows
an intriguing arc-like structure 1.9~arcsec from the nucleus, which
may be the remnant of a past galaxy merger. 1959+650 is found to be
hosted by a gas rich galaxy, with a total dust mass of roughly
$5\times10^5$ M$_\odot$. Large quantities of dust are quite often
observed in elliptical galaxies, but perhaps due to the difficulty of
observing them near a bright nucleus, have not been reported for BL
Lac hosts. The weakness of the emission lines in this class of AGN is
sometimes attributed to a lack of gas surrounding the central power
source (but see Scarpa \& Falomo 1997) but the present observations of
1959+650 certainly argue against such an idea, demonstrating that at least 1
BL Lac resides in a gas rich galaxy.

\acknowledgements
Support for this work was provided by NASA through grant 
GO06363.01-95A from the Space Telescope Science
Institute, which is operated by AURA, Inc., under NASA contract
NAS~5-26555. We thank Eric Perlman for providing useful information.

\begin{deluxetable}{llllrc}
\tablenum{1}
\tablecaption{Journal of the Observations}
\tablehead{
 \colhead{object}   & \colhead{$\alpha (2000)$} &
\colhead{$\delta (2000)$} &\colhead{date} & \colhead{exp} & \colhead{z$^{(a)}$}
}
\startdata
$0033+595$A & 00:35:52.549 &~~59:50:03.47& ~3 Mar 96 &  1060 & ...      \nl
$0033+595$B & 00:35:52.734 &~~59:50:04.18&           &       &          \nl
$0138-097$  & 01:41:25.76  &--09:28:43.4 & 28 Sep 96 &   840 & 0.733 (1)\nl
$0502+675$  & 05:07:56.25  &~~67:37:24.4 & ~2 Mar 96 &   740 & 0.314 (2)\nl
$0806+524$  & 08:09:49.15  &~~52:18:58.7 & 11 Sep 96 &   610 & 0.136 (2)\nl
$1440+122$  & 14:42:48.35  &~~12:00:40.5 & ~7 May 97 &   320 & 0.162?(3) \nl
$1517+656$  & 15:17:47.60  &~~65:25:23.9 & ~1 Feb 97 &   614 &$>$0.7 (4)\nl
$1959+650$  & 19:59:59.87  &~~65:08:54.1 & ~9 Jan 97 &   302 & 0.048 (5)\nl
\enddata
\tablenotetext{a}{References for redshifts are: (1) Stocke \& Rector 1997;
(2) Perlmann, private communication; (3)Schachter et al. 1993 (see text for more on
this value); (4) Beckmann et al.,
private communication; (5) Perlmann et al. 1996.}
\end{deluxetable}

\begin{deluxetable}{lccc}
\tablenum{2}
\tablewidth{4in}
\tablecaption{Companion objects near 0138--097}
\tablehead{
 \colhead{object} & \colhead{Separation} & \colhead{PA$^{(a)}$} & \colhead{$m_R$}\\
 &  \colhead{(arcsec)} & \colhead{(degrees)}
}
\startdata
0138--097&$\dots $&$\dots $&$17.48\pm 0.05$\nl
B       & 2.32   & 175    &$ 23.2 \pm 0.3 $\nl
C       & 1.45   & 198    &$ 22.1 \pm 0.1 $\nl
D       & 1.92   & 303    &$ 24.1 \pm 0.4 $\nl
E       & 4.87   & 244    &$ 24.6 \pm 0.3 $\nl
\enddata
\tablenotetext{a}{Position angle from north toward east.}
\end{deluxetable}

\newpage

{\bf Figure Captions}

\noindent
{\bf Fig. 1}
Central part of the HST WFPC2 F702W image of 0033+595 (PC camera).
The proposed counterpart of this BL Lac object is resolved into two
point sources, ``A'' and ``B,'' which have comparable brightness
(``A'' is slightly brighter than ``B'' in R but the situation is
reversed in U) and are separated by 1.58~arcsec. It is reasonably
probable that we would find one such pair, simply by chance, in 
a set of 100 observations. A faint galaxy (``G'')
is also detected just to the East of the two point sources. The arrow
indicates north and is 1.84~arcsec (40 pixels) long.

\noindent
{\bf Fig. 2}
Radial profiles of 0033+595 ``A'' (open squares) and ``B'' (crosses), 
and a PSF model including large-angle scattered light (solid line). 
The radial profile of ``B'' has been shifted by 0.2 magnitudes
to match that of ``A'' in order to better 
emphasize differences at large radii. Within the uncertainties, 
both appear unresolved.

\noindent
{\bf Fig. 3}
Finding chart for 0033+595. This image was obtained with the
Canada-France-Hawaii telescope in the I band. Object
 ``B'' is marked as the most probable identification of the BL Lac object. 
North in on top, east on left.

\noindent
{\bf Fig. 4} HST WFPC2 and NICMOS images of the BL Lac object
0502+675. {\bf Left:} Central part of the PC F702W image, showing two
objects separated by 0.33~arcsec. Here the image is printed with a
gray-scale emphasizing the brightest structures, so that the galaxy is not
visible. The light surrounding the brighter source is due to the bright
wings of the PSF. No other sources are detected within a radius of
almost 10~arcsec. Figure orientation (north is roughly at the
bottom) and scale can be derived from the position angle and separation
of the fainter source with respect to the brighter
(P.A.=$24^\circ$ from north toward east and separation 0.33~arcsec).
{\bf Center:} Central part of
the NICMOS F160W image, with same linear scale and orientation as in 
the previous panel, at approximately the same intensity stretch. With this
particular intensity stretch, all visible structures surrounding 
the brighter source (apart from the companion) are from the PSF.
{\bf Right:} Same NICMOS field as in (b), but with the two point
sources subtracted, and intensity stretched to bring out the galaxy.

\noindent
{\bf Fig. 5}
Comparison of the radial profiles of the brighter (``A,'' squares)
and fainter (``B,'' crosses) point sources that comprise 0502+675.
The solid line represents the HST PSF, which includes
scattered light at large radii (see text for details). 
To better campare the two point sources, the 
radial profile of object ``B'' has been shifted up by
1.4~magnitudes to match the few innermost pixels of the brightest
source profile.

\noindent
{\bf Fig. 6}
The complex field of the BL Lac object 1440+122, from the central
part of the WFPC2 F702W PC image.
The BL Lac is an elliptical galaxy with a bright point source 
at its center (``A''), and there is a second point source (``B'') 
0.29~arcsec to the east. A second galaxy, of similar size and brightness, 
lies 2.53~arcsec to the west; such luminous
companions are not commonly found near BL Lac objects.
The probability is low that the close pair ``A'' and ``B'' represents a
chance superposition of two unrelated objects.
The arrow points north and is 1.38~arcsec (30 pixels) long.

\noindent
{\bf Fig. 7}
The radial profile of the BL Lac object 1440+122 (``A'' in Fig.~6),
described using a two-component model consisting of
a de Vaucouleurs law (dashed line) plus a point source 
(dotted line). The sum of the two components (solid line) follows well
the observed profile (squares) except within the first 0.1~arcsec where
the PSF is undersampled.

\noindent
{\bf Fig. 8}
Part of the WFPC2 F720W PC image of the BL Lac object 1517+656,
after smoothing with a Gaussian of $\sigma=0.8$ pixels. 
Three narrow arclets surround the central point source, at position angles 
2$^\circ$, 145$^\circ$, and 275$^\circ$, describing an almost
perfect ring.
Two bright spots are also resolved, at position angles
126$^\circ$ and 260$^\circ$. 
The arrow points north and is 1.38~arcsec (30 pixels) long.

\noindent
{\bf Fig. 9}
Part of the WFPC2 F702W PC image of the high redshift
BL Lac object 0138--097, after Gaussian smoothing with 
$\sigma = 2$~pixels.
An unusual number of close companion galaxies can be seen clearly 
(B-D are named following Heidt et al. 1996). 
The arrow points north and is 1.84~arcsec (40 pixels) long.

\noindent
{\bf Fig. 10}
The average radial profile of 0138--097 (squares) compared with the PSF 
profile (solid line).
There is some departure from the PSF but we can not claim
the object is resolved (see text).

\noindent
{\bf Fig. 11}
Part of the WFPC2 F702W PC image of the BL Lac object 0806+524, 
showing a well resolved host galaxy surrounding a bright central point source.
An unusual arc-like structure is detected 1.9~arcsec south of the nucleus,
possibly a bright elliptical shell or the remnant of a previous
gravitational interaction.
The arrow points north and is 1.38~arcsec (30 pixels) long.

\noindent
{\bf Fig. 12}
The WFPC2 F720W image of 1959+650 after subtracting a scaled PSF. 
There is a large dust lane (pale arc) 0.8~arcsec north of the nucleus,
an unusual feature in BL Lac host galaxies. The straight structure
inclined at $\sim 45^\circ$ is a remmant spike from the PSF.
The arrow points north and is 1.38~arcsec (30 pixels) long.

\newpage

\begin{figure}
\psfig{file=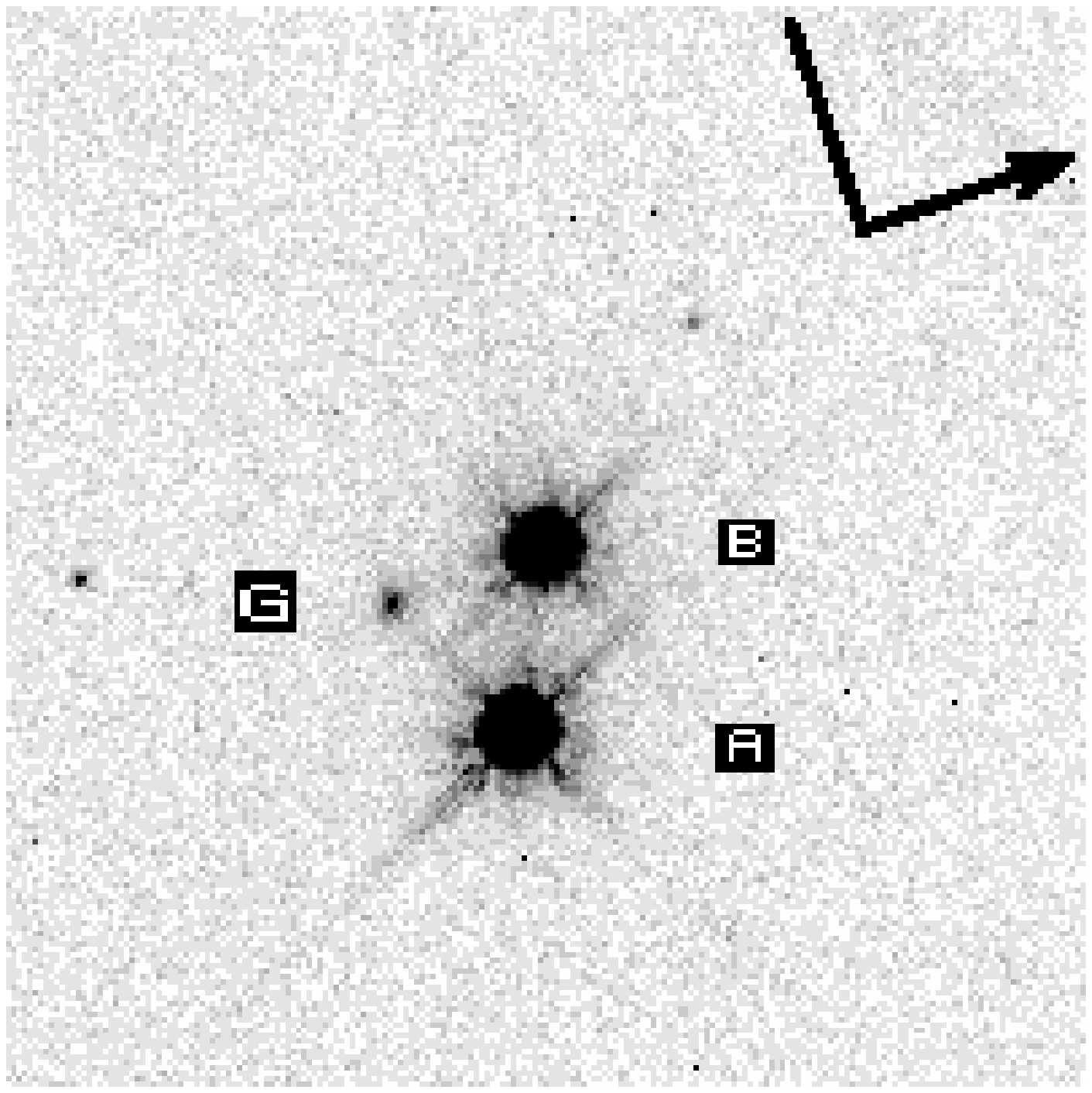,width=10cm}
\caption{\label{fig1}} 
\end{figure}
 
\begin{figure}
\psfig{file=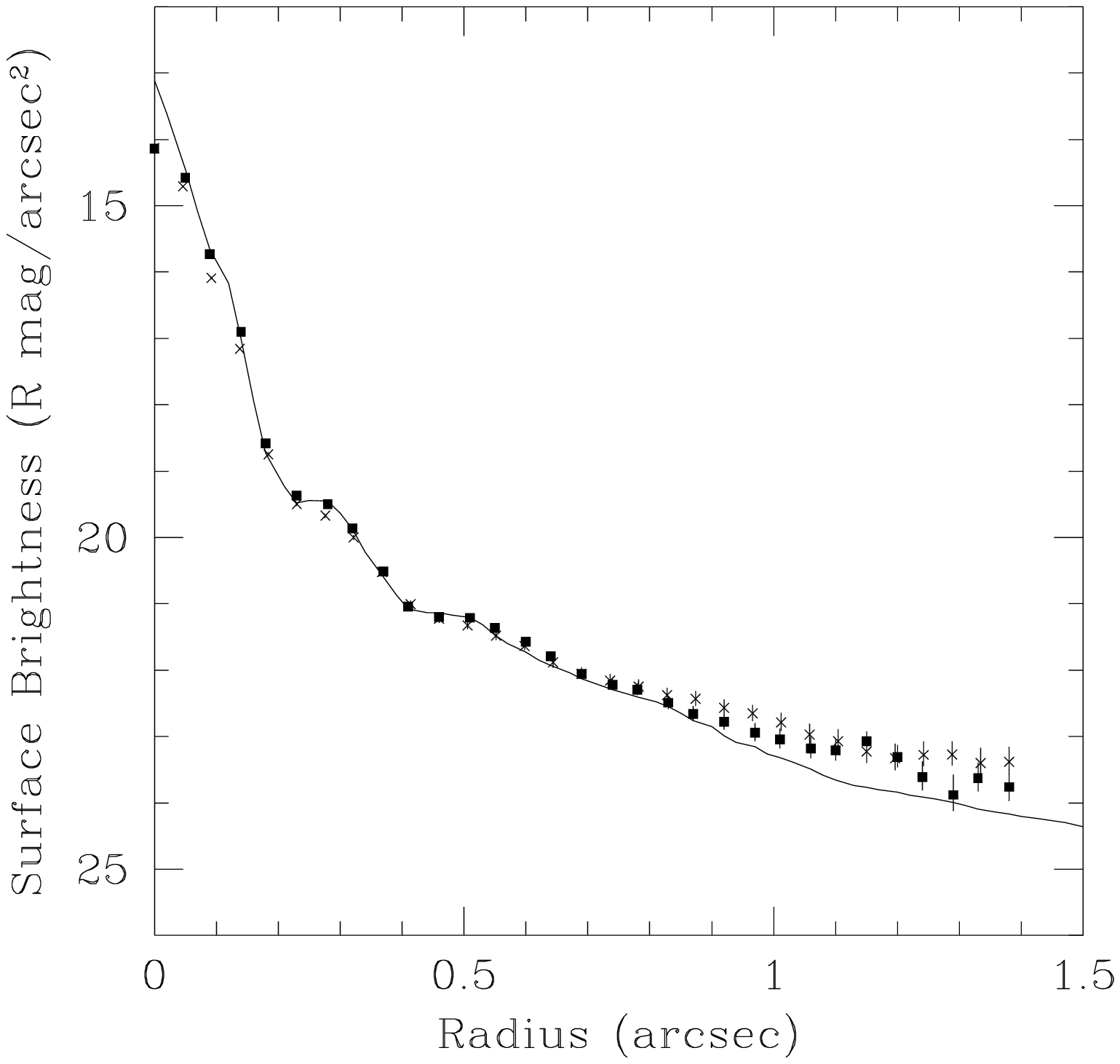,width=14cm}
\caption{\label{fig2}} 
\end{figure}
 
\begin{figure}
\psfig{file=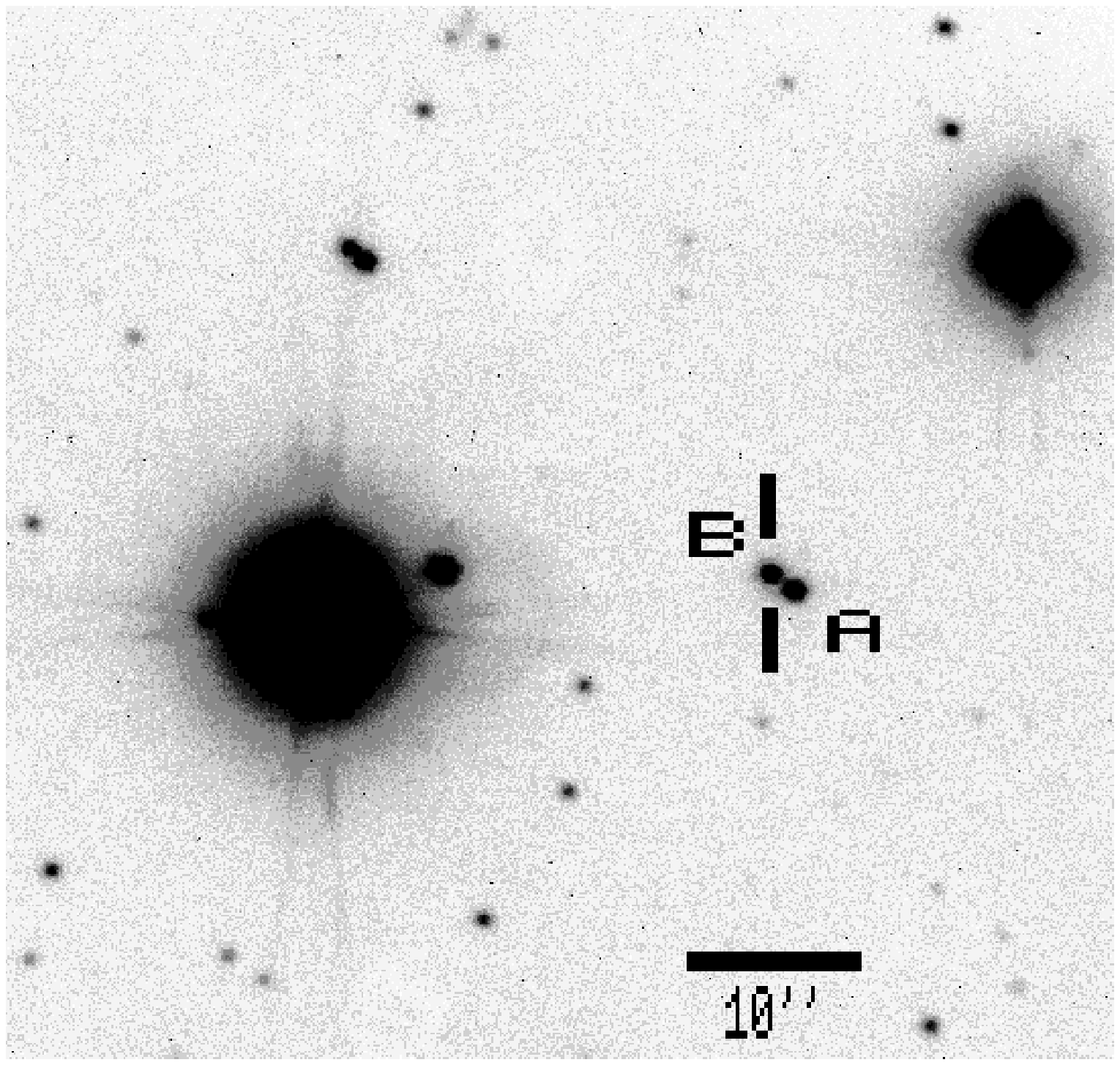,width=12cm}
\caption{\label{fig3}} 
\end{figure}

\begin{figure}
\centerline{
\psfig{file=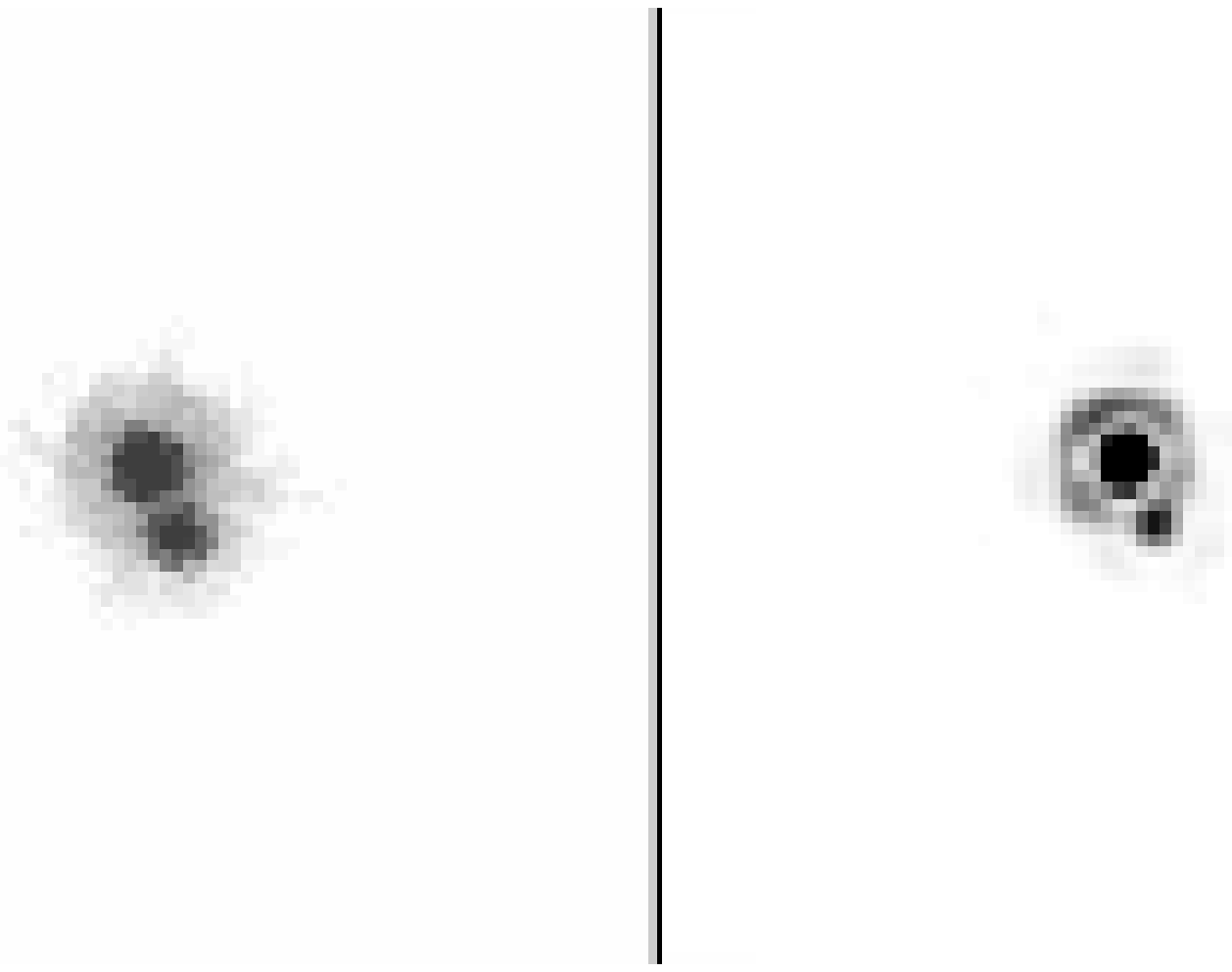,width=0.5\linewidth} 
\psfig{file=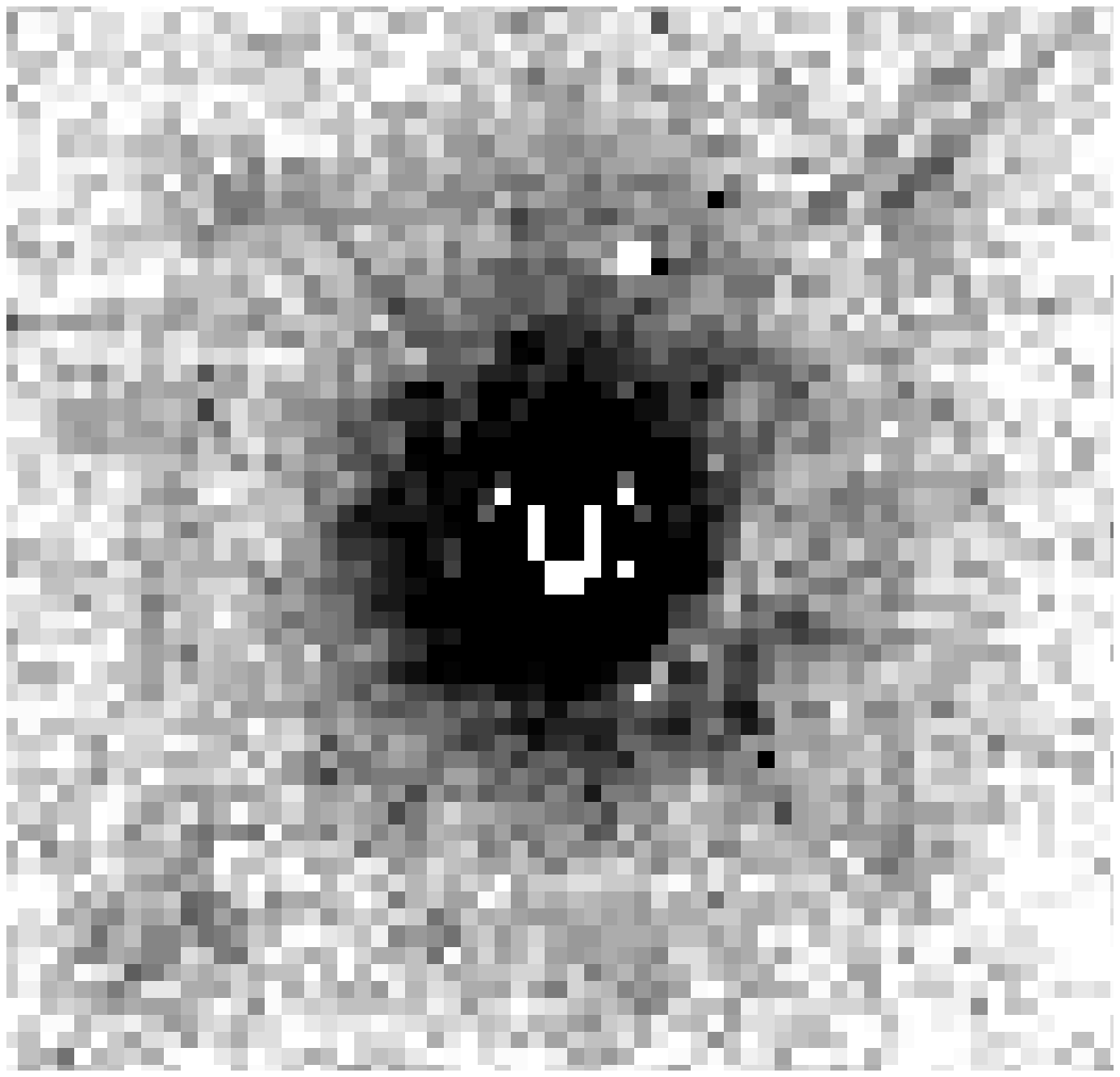,width=0.28\linewidth}
}
\caption{\label{fig4}} 
\end{figure}
 
\begin{figure}
\psfig{file=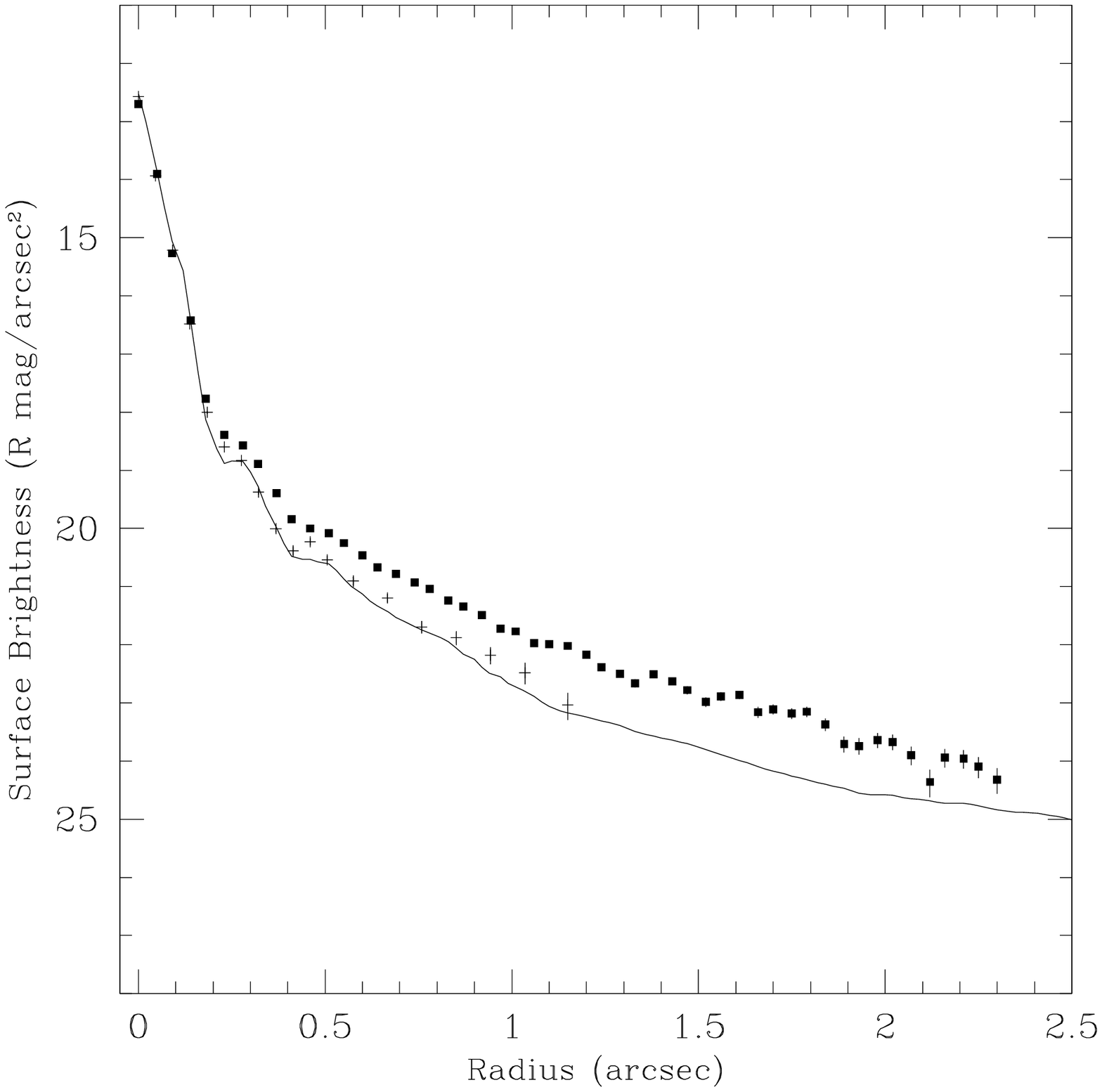,width=10cm}
\caption{\label{fig5}}
\end{figure}
 
\begin{figure}
\psfig{file=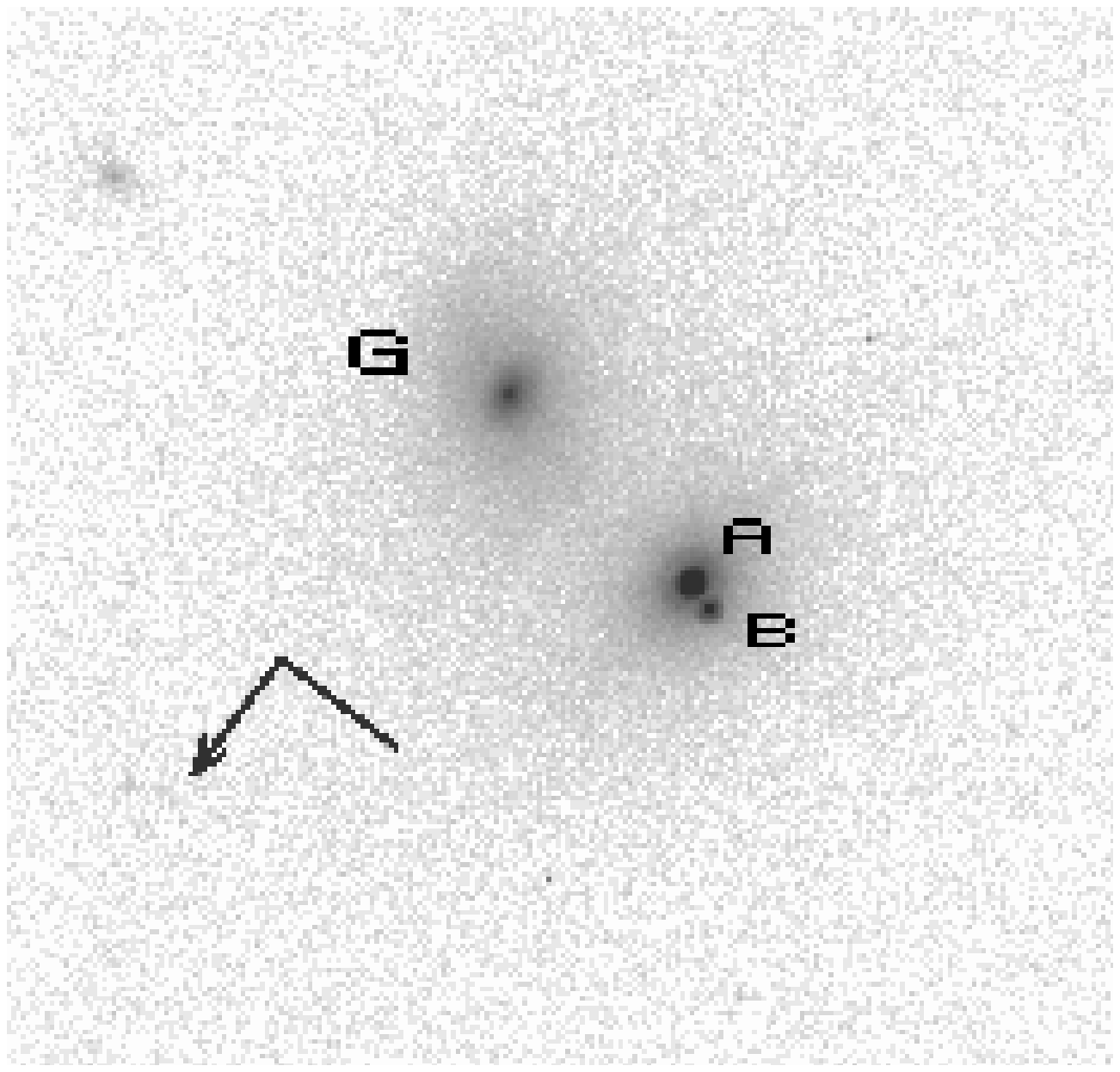,width=10cm}
\caption{\label{fig6}}
\end{figure}
 
\begin{figure}
\psfig{file=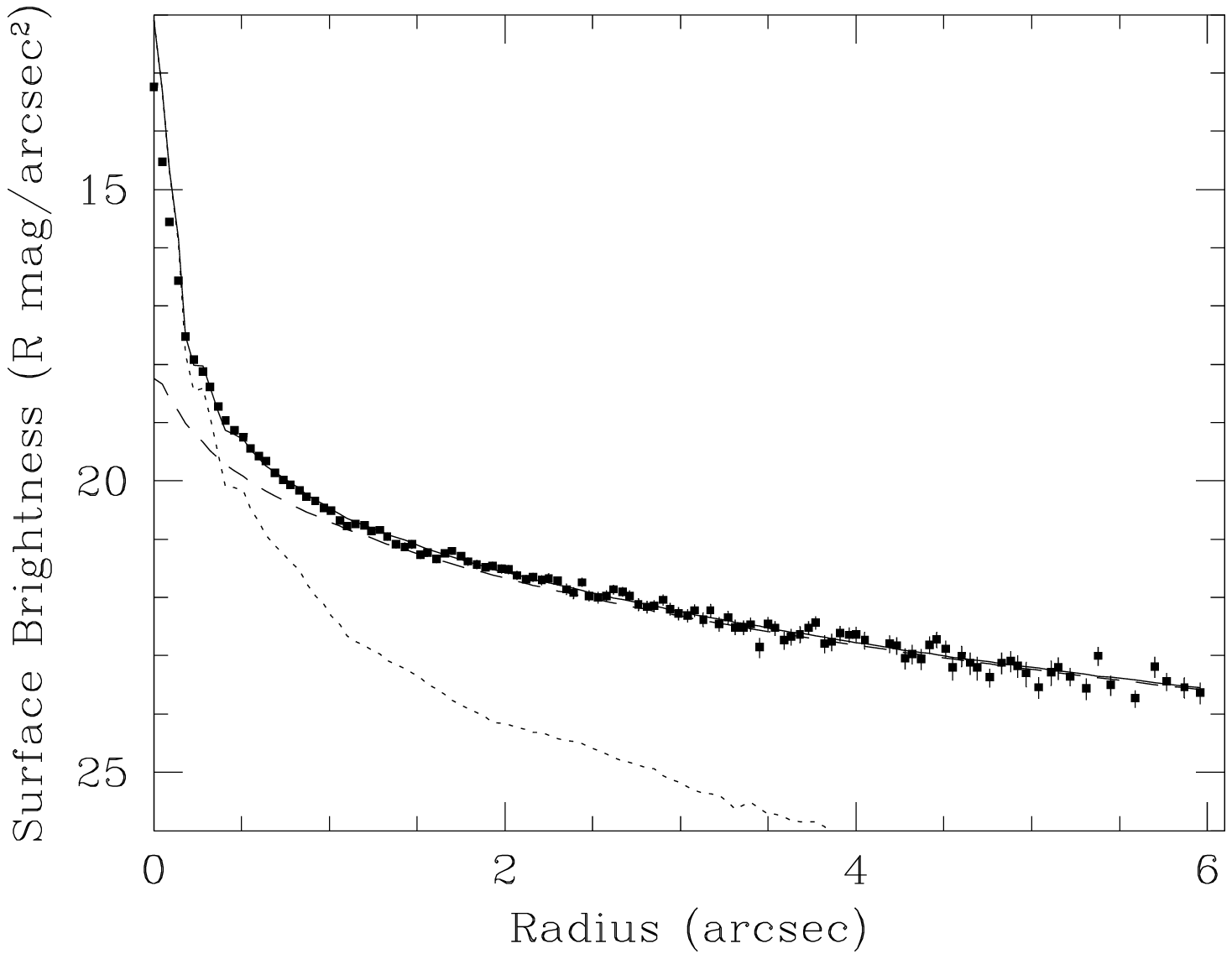,width=15cm}
\caption{\label{fig7}}
\end{figure}
 
\begin{figure}
\psfig{file=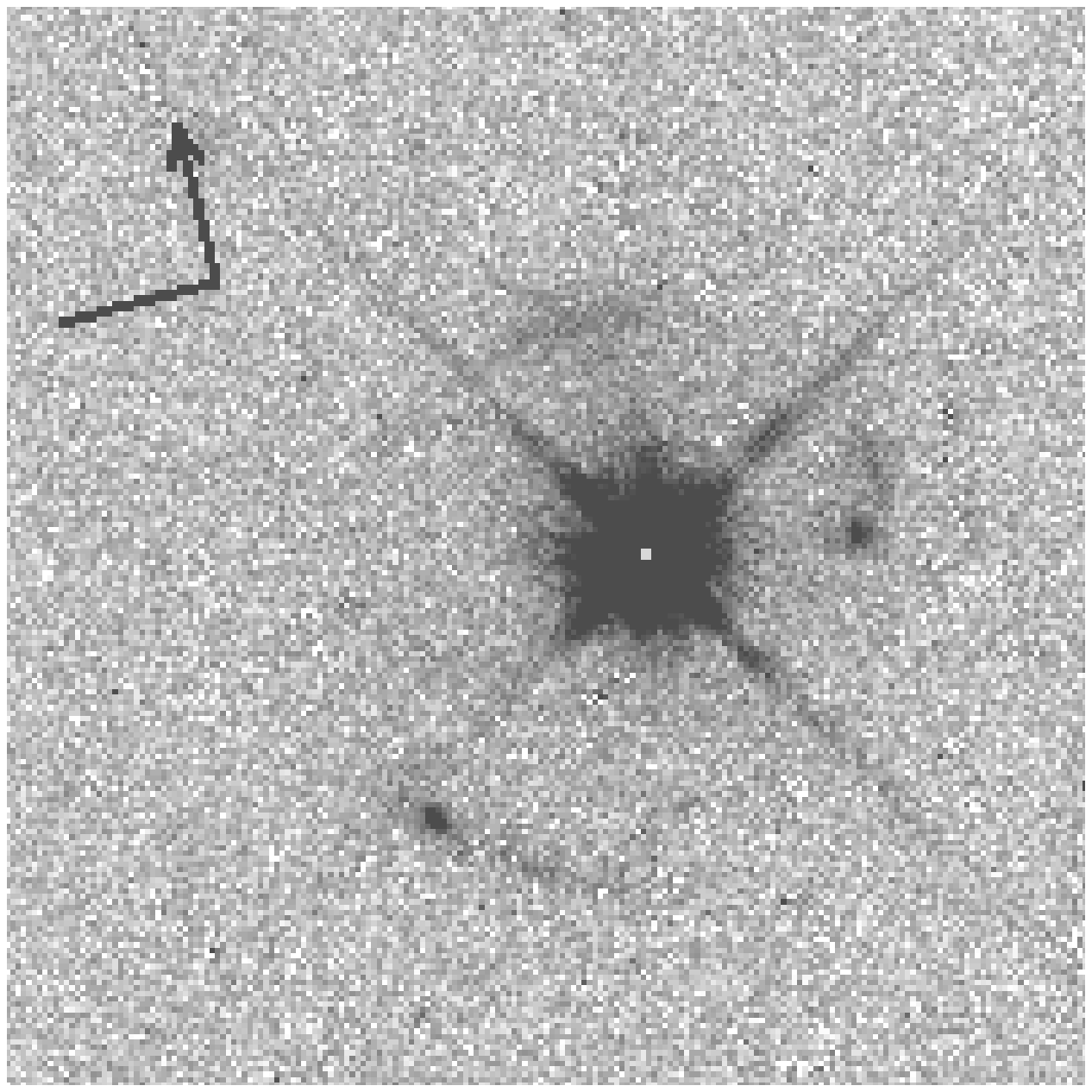,width=12cm}
\caption{\label{fig8}} 
\end{figure}
 
\begin{figure}
\psfig{file=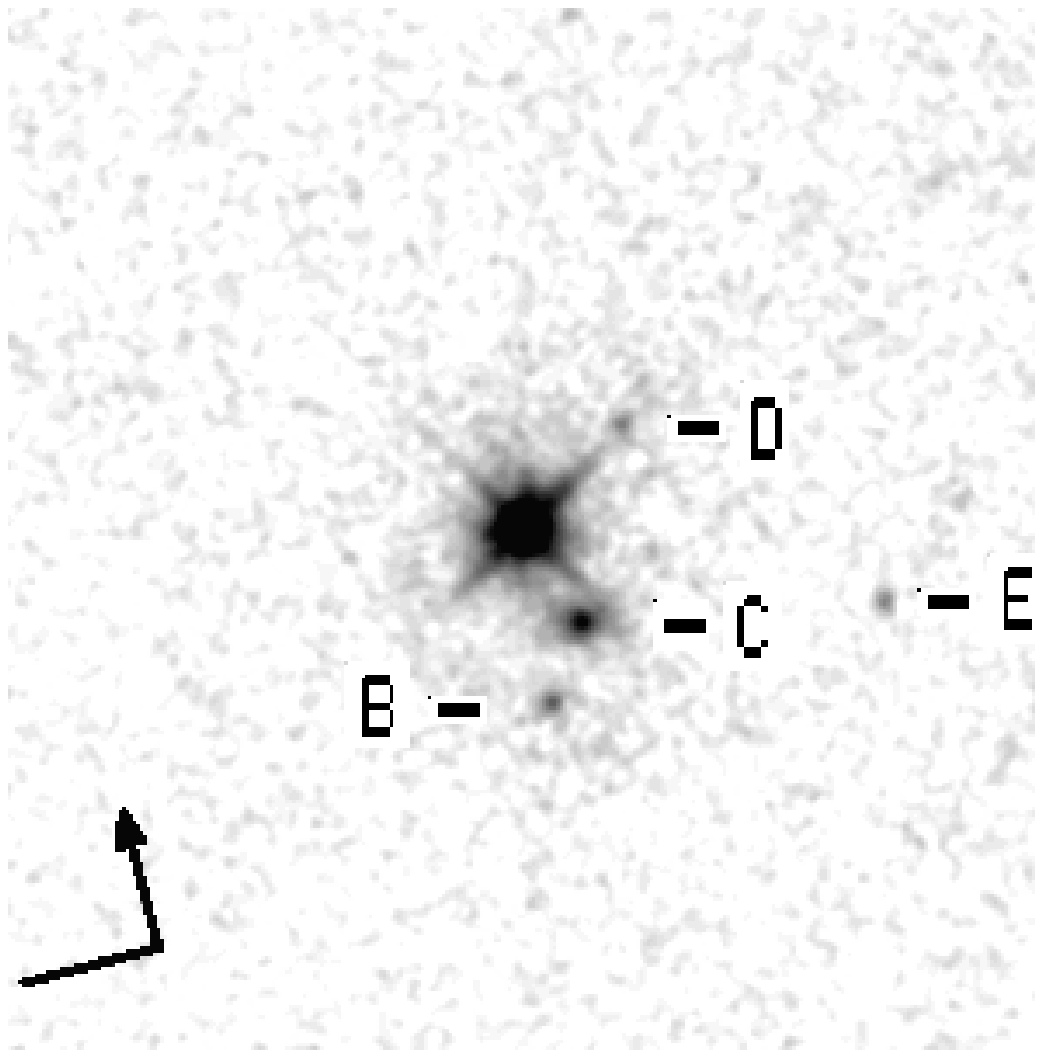,width=8cm}
\caption{\label{fig9}} 
\end{figure}
 
\begin{figure}
\psfig{file=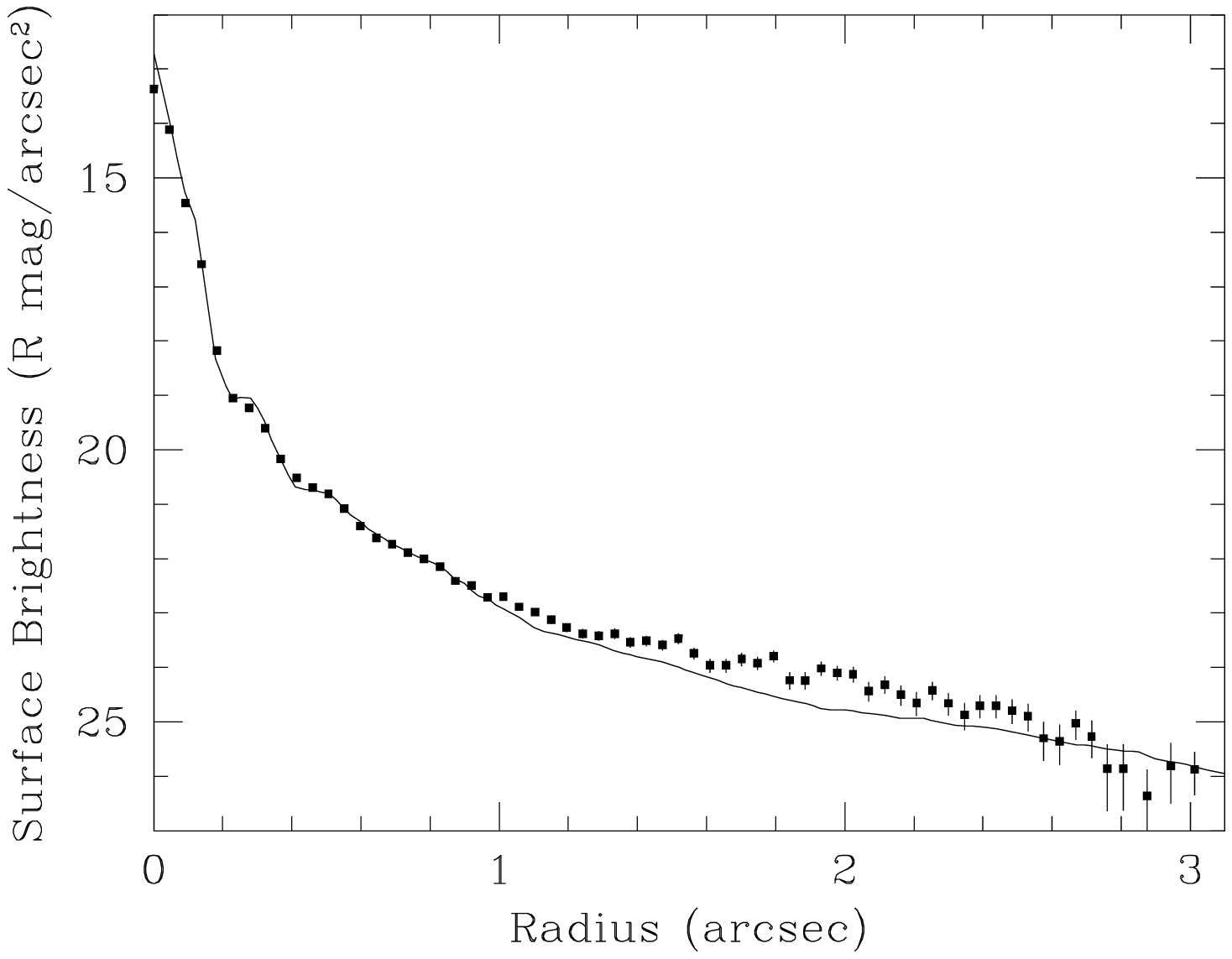,width=12cm}
\caption{\label{fig10}}
\end{figure}
 
\begin{figure}
\psfig{file=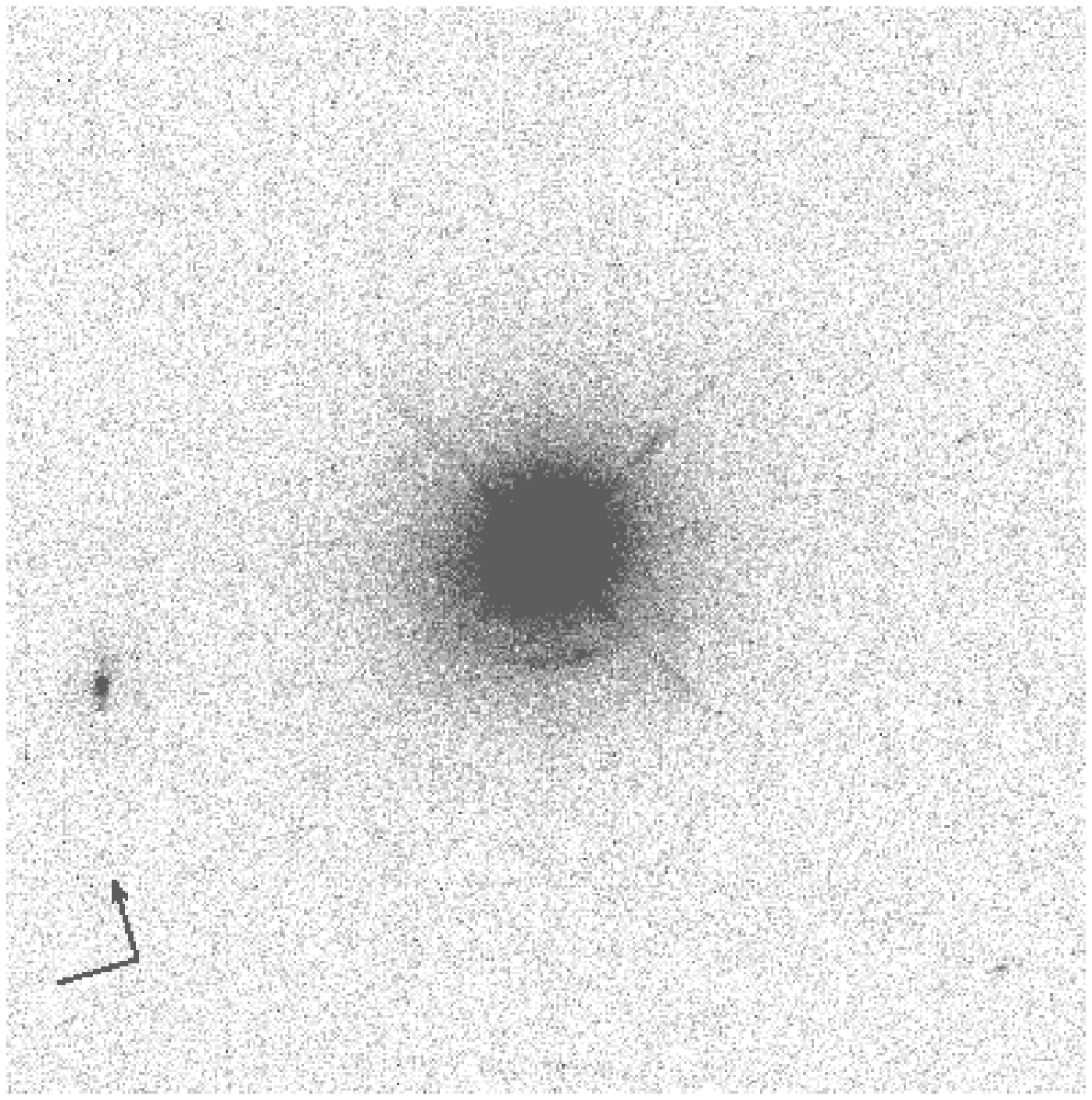,width=10cm}
\caption{\label{fig11}} 
\end{figure}
 
\begin{figure}
\psfig{file=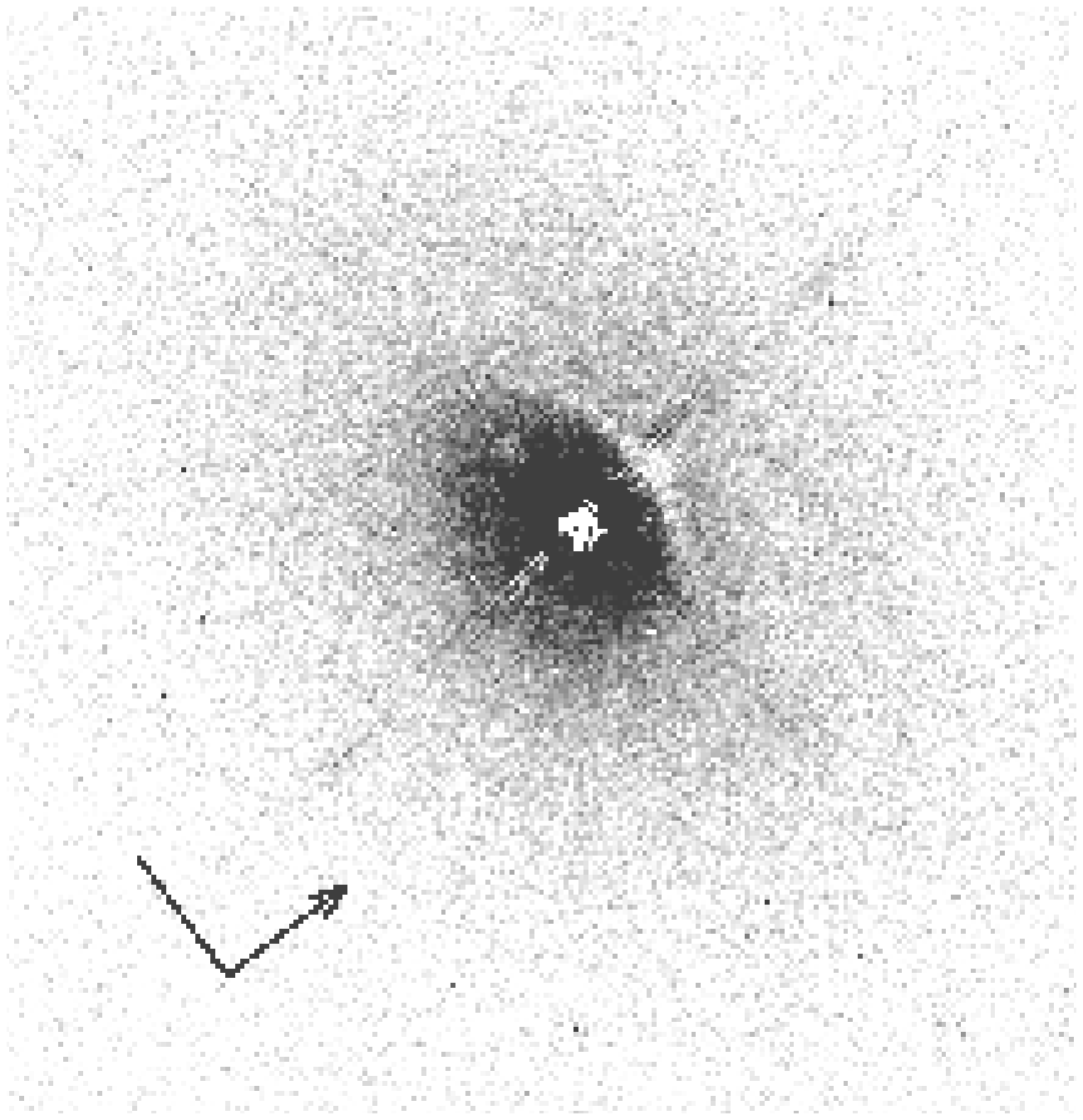,width=10cm}
\caption{\label{fig12}} 
\end{figure}

\end{document}